# In-plane subwavelength near field optical capsule for lab-on-a-chip optical nano-tweezer

OLEG V. MININ,[1] SHUO-CHIH CHIEN[2], WEI-YU CHEN,[2] CHENG-YANG LIU,[2,**] IGOR V. MININ,[1,*]

[1]Tomsk Polytechnic University, Tomsk, 634050, Russia
[2]Department of Biomedical Engineering, National Yang-Ming University, Taipei City, Taiwan
*Corresponding author: cyliu66@ym.edu.tw, prof.minin@gmail.com

**In this letter, we propose a new proof-of-concept of optical nano-tweezer on the basis of a pair of dielectric rectangular rods capable of generating a novel class of controlled finite-volume near field light capsules. The finite-difference time-domain simulations of light spatial structure and optical trapping forces of the gold nanoparticle immersed in water demonstrate the physical concept of an in-plane subwavelength optical capsule, integrated with the microfluidic mesoscale device. It is shown that refractive index and distance between dielectric rectangular rods can control the shape and axial position of the optical capsule. Such an in-plane wavelength-scaled structure provides a new path for manipulating absorbing nano-particles including bio-particles in a compact planar architecture and should thus open promising perspectives in lab-on-a-chip domains.**

http://dx.doi.org/

The capture of particles with a refractive index lower than the refractive index of the environment, or particles, whose surface is highly absorbing, is performed at a minimum intensity [1]. Consequently, beams designed to capture such particles should have local intensity minima surrounded by intensity maxima. Such beams, the longitudinal intensity distribution of which in the waist region has an isolated minimum of intensity, uniformly surrounded by zones of maximum light energy, and the cross section has the form of a light ring with a dark center, are called "bottle" beams [2], or sometimes-called light capsules.

To create bottle beams, various methods are used, such as generation of Bessel beams based on axicons [3], interference of Gaussian beams with different waist [4], when destructive interference leads to the formation of a region with a minimum intensity at mutual focusing of two beams, and based on the superposition of different modes [5,6]. Some methods use coherent superposition of two vortex beams [7], Gaussian beams [8], to name a few.

However, these methods for generating optical capsules either require precise tuning of the optical system or are not universally applicable. Moreover, the known optical bottle beams either have axial symmetry or are three-dimensional, which limits their use in integrated compact devices of the "laboratory on a chip" type, and the particle confinement region is located at distances much longer than the wavelength of the radiation used. Known optical capsules are far-field systems and therefore obey the fundamental diffraction limit. This limits their potential.

Transparent dielectric mesoscale particles are a class of modern micro-optics with elegant focusing of light and low losses [9]. In classical optical tweezers, in the manipulation of Rayleigh particles one of the advances based on reducing the trap volume, based on photonic jet effect, which increases the optical trapping force. Optical traps based on a photonic jet lead both to an increase in the thermal stability of the trap and to an increase in the intensity of radiation generated by optically captured nanoparticles [10-15]. In 2012 it was shown, that under the Bessel-Gauss beam and an aberrated focused Gaussian beam particular illumination of dielectric microsphere the optical-bottle-beam-like photonic najojets is formed [16]. Recently [17] it was considered the formation of a three-dimensional bottle beam by nanostructuring the shadow surface of a microsphere using diffraction gratings offset from each other [2].

However, all of the above methods of generating optical capsules cannot be applied for use, for example, in a planar configuration for lab-on-a-chip devices at wavelength-scaled dimensions. At the same time, the division of the incident wave beam into independent partial beams with their subsequent interference to create localized regions of the required shape is possible using a pair of dielectric rods. It was shown in [18] that dielectric structures based on a pair of dielectric rectangular rods with the same refractive index are capable of focusing radiation at a focal length that depends on the distance between dielectric particles. We recently proposed [19] and demonstrated structures with a pair of dielectric rectangular rods capable of generating diffraction limited photonic hooks at wavelength-scaled distances from their output surface when illuminated by a linearly polarized plane wave. It has been shown how such curved light beams can be produced by correctly choosing the size and refractive index of two dielectrics, so that one of them is capable of generating high scattering and the other low diffraction along the propagation axis.

Inspired by these recent discoveries, as well as the interesting possibilities that optical capsules have to offer in a variety of applications, including sensors and optical capture, in this work we propose and demonstrate the possibility of creating in-plane structured light beams such as optical capsules using a pair dielectric rectangular particles with different sizes and/or refractive index. An attractive feature of this design is its extreme simplicity, compactness (using rectangular shapes) and the ability to dynamically control the parameters of the region of field localization, for example, by measuring the distance between structures, making it possible to overcome the diffraction limit by creating a planar optical trap of subwavelength size and increased strength.

Figure 1 illustrates the schematic stereogram of the proposed microfluidic device with a pair of dielectric rectangular structures. The dielectric rectangular structures are placed in a microchannel and the material of the microfluidic device is polydimethylsiloxane (PDMS). The length and width of the dielectric rectangular structures are $h = 0.3$ μm and $l = 0.9$ μm. The refractive index of the rectangular structures is $n_1$. The fluid flow in the microchannel is water ($n_2 = 1.33$). The pair spacing between the two dielectric rectangular structures is $d$. The laser beam with λ=532-nm wavelength propagates through a pair of dielectric rectangular structures in the positive $y$ direction. The high-intensity optical bottle beam is generated from the shadow-side surface of the dielectric rectangular structures. The spatial intensity distributions acquired in the $x$-$y$ plane are examined in order to determine the spatial distribution properties of localized in-plane optical bottle beam. The distance between the rectangle surface and the center of the null field in the optical capsule along the $y$ axis is $f$. It is noteworthy that in the proposed configuration, the optical capsule is created along the direction of flow, in this case - liquid, and not across, as in all configurations known to the authors, due to the "hole" between a pair of dielectric rods.

The finite-difference time-domain (FDTD) method [20] is performed to observe the spatial intensity distributions generated by a pair of dielectric rectangular structures. The light source was a linearly polarized plane wave. The boundary conditions in the FDTD modeling domain were set as perfectly matched layers and the cell size was λ/90 throughout the modeling domain. The dependences of spatial intensity distributions on the refractive index contrast and the pair spacing of rectangular structures are investigated in detail.

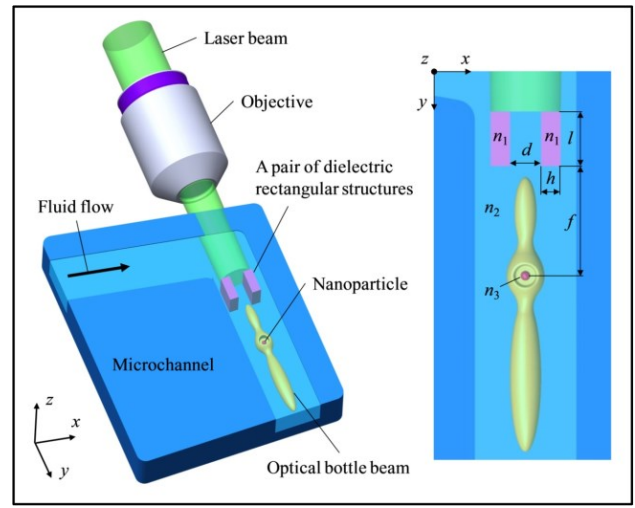

Fig. 1. Schematic stereogram of a microfluidic device with a pair of dielectric rectangular structures. The insert indicates the definitions of geometric and material parameters.

Figure 2 illustrates the conditions for the formation of an in-plane optical bottle using a pair of dielectric rods with a fixed distance between them and different optical contrast. At a low optical contrast (K = $n_1/n_2$ = 1.65/1.33 = 1.24), the region of radiation localization is inside between the rods and does not go beyond the structure.

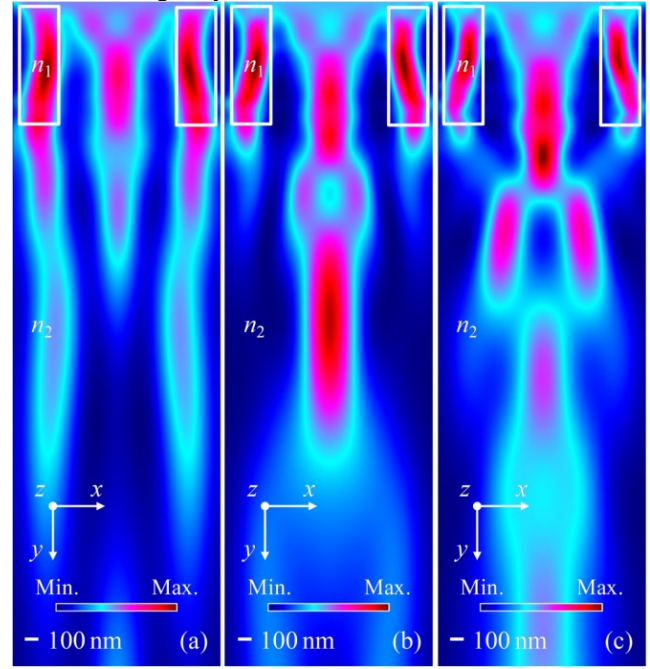

Fig. 2. Calculated electric field intensity distributions generated by a pair of dielectric rectangular structures in water at (a) $n_1 = 1.65$, (b) $n_1 = 1.85$, and (c) $n_1 = 2$. The pair spacing is $d = 0.9$ μm.

With an increase in contrast to K = 1.39 (by 12%), the region of radiation localization takes on a form characteristic of an optical capsule, and this region is displaced beyond the boundaries of the dielectric rods. The optical barrier is harmonic in the Y-direction of propagation of radiation, while in the transverse directions it has an approximately elliptical shape and maxima separated by the width of the main central lobe. Obviously, the optical barrier before or after the focal point arises due to the interference of converging or diverging localized beams generated by a pair of dielectric rods - due to the superposition of photonic flows from a pair of rods and an optical wave passing through the central transparent hole between them. With a further increase in the contrast to K = 1.5, the closed region of localization of radiation from minima inside is violated, forming a central maximum and two lateral radiation maxima characteristic of an ordinary lens.

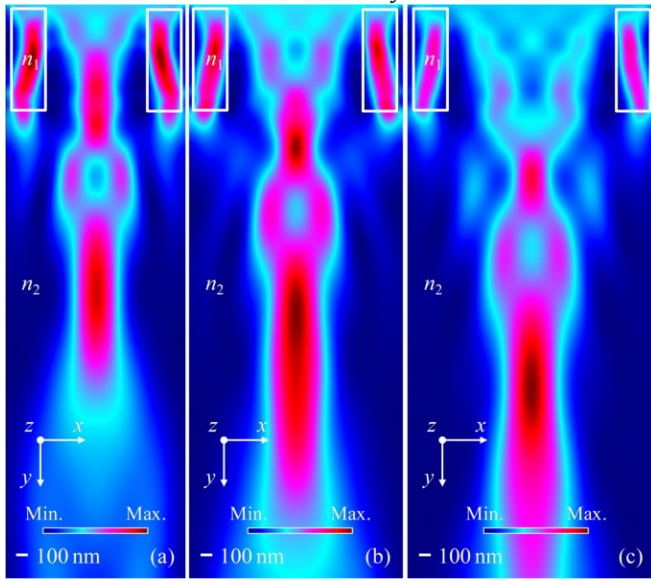

Fig. 3. Calculated intensity distributions generated by a pair of dielectric rectangular structures in water at (a) $d$ = 0.9 μm, (b) $d$ = 1.2 μm, and (c) $d$ = 1.5 μm. The refractive index of the rectangular structures is $n_1$ = 1.88 and the incident wavelength is 532 nm.

It is noteworthy that, based on our method, not only the size, but also the position in space (along the direction of radiation propagation) of the light capsules can be continuously controlled by adjusting one parameter. The conditions for the formation of an in-plane optical bottle of various geometric shapes using a pair of dielectric rods with a fixed optical contrast are shown in Fig. 3, illustrating the dependence of the size, shape and position in space of the optical capsule on the distance between the dielectric rods.

It is obvious that light capsules can have different geometries due to different choices of the parameters of the dielectric structure. Thus, we find that not only the size, but also the shape (the degree of closure of the optical bottle) changes with the change in the distance between the rods. As the distance between the rods increases, the focal length of the optical capsule also increases - the distance from the shadow surface of the rods to the center of the area with the minimum intensity. The most closed area of increased intensity is observed at a relative distance between the rods equal to $d_1 = d/\lambda n_2 = 1.696$ ($d$ = 1.2 μm).

Thus, if the parameters of the dielectric rods are chosen correctly, the characteristics of the formed flat bottle beam can be effectively modulated. Therefore, this means that with our method it is possible to obtain a controlled light capsule that almost ideally surrounds an area of minimum intensity.

The field intensity distribution in the focal area is a two-dimensional (2D) optical barrier for the formation of an optical capsule, shown in Fig.4a. An almost perfect closed area of low intensity surrounded by areas of high intensity is clearly visible. It is noteworthy that this can be easily achieved by choosing the parameters of the dielectric rods.

To quantitatively analyze the result, we extract and plot the field intensity along the y and x axes in Fig. 4 (c). As shown in Fig. 4, we define the transverse size ($b_x$) and longitudinal size ($b_y$) of the minimum intensity field in the optical capsule as the distance between two intensity maxima in the transverse and longitudinal directions, respectively. The Rayleigh range of two focal regions of increased intensity, forming an almost closed region with a minimum intensity inside, is shown in Fig. 4b and ranges from x = 0.89 um to x = 0.78 um. It follows from Fig. 4b that the transverse size of the trap is subwavelength (less $\lambda/2n_2$) when the refractive index of the rods is less than 1.75. The longitudinal amplitude curve shows a typical saddle shape that has a low intensity region with a longitudinal distance of 1.67 um. Here, the longitudinal distance is defined as the interval between the peaks of the front and back optical barriers near the longitudinal dark region.

The dependence of the focal length on the optical contrast is shown in Fig. 4c. It follows from these figures that the transverse dimension of the light capsule is subwavelength when the optical contrast of a pair of dielectric rods is less than $n_1/n_2 < 1.65$. This area is shown with a dashed line in Figure 4b.

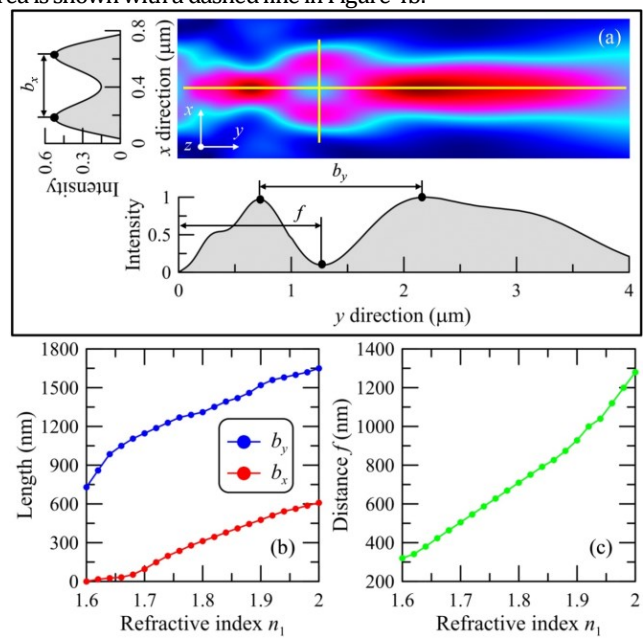

Fig. 4. (a) Sketch definition of the axial and radial lengths of the optical capsule. The yellow lines indicate the propagation direction ($y$ axis) and the transverse direction ($x$ axis). (b) The relationship between the axial ($b_y$) and transverse ($b_x$) lengths of the mininal field intensity area vs refractive index $n_1$. (c) The location of the optical capsule ($f$) along y-axis as a function of the refractive index $n_1$.

The characteristics of the in-plane evolution of the light capsule correspond to the characteristics of a typical beam with an optical bottle [2]. Obviously, as the wave propagates, the central peak intensity of the light field gradually decreases and a dark focus with a low intensity appears until a sharp flat ring is formed. After passing through the position y = 1.25 um, a reverse diffraction process occurs. Namely, the edge of the ring of light gradually becomes ambiguous, and the central dark focus begins to shrink. Finally, the intensity of the light wave is concentrated in the center and again becomes a peak at y = 2.2 um.

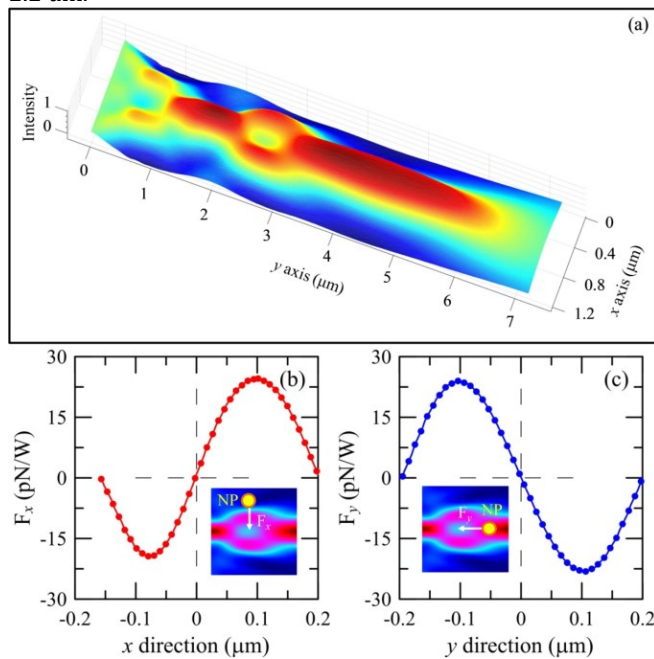

Fig. 5. (a) Three-dimensional intensity model of the optical capsule formed by a pair of dielectric rectangular structures in water. Optical forces of a 100 nm gold nanoparticle as functions of (b) *x* and (c) *y* directions.

Finally, Figure 5 demonstrates the capture of a nanoparticle in an optical capsule. The optical forces acting on the nanoparticle were calculated in accordance with the results of Refs. [21,22]. For 100 nm gold (Au) nanoparticles in water, the position of the plasmon resonance is about 580 nm [21]. Thus, at a wavelength of λ=532 nm, where absorption cross section of Au nanoparticle has maximum, a blue detuning is observed with respect to the plasmon resonance of Au nanoparticles. In [21] it was shown that in the process of interaction of Au nanoparticle with radiation near the plasmon resonance, the scattering force $F_{scat}$ dominates in comparison with gradient force $F_{grad}$, and thus gold nanoparticles are attracted by regions of low light intensity.

The power of the laser for calculation of optical force was 10 mW. As it can be clearly seen from Fig.5 that the gold single-nanoparticle with 100-nm diameter and with the refractive index of $n_3$ = 0.5439 + 2.231i can be trapped by the proposed optical capsule, which is of considerable interest, for example, for *in-vivo* cell capture experiments [23]. It is seen that the optical forces (both $F_x$ and $F_y$) changes sign (from attraction to repulsion) around the center of the region with low field intensity, and creates restorative forces directed towards the center of the optical capsule to serve as an optical barrier to isolate the nanoparticle from being disturbed.

In conclusion, we observed effects of the in-plane near field optical capsule formation by the pair of dielectric rectangle rods. The two dimensional optical capsule generation considered in this Letter does not require the structuration of illuminated beam. In contrast to known methods of the bottle beam generating, the optical capsule parameters can be simply controlled by refractive index and distance between pair of dielectric rods. The future research can explore the influence of the pulsed laser illumination on the optical capsule parameters.

**Funding.** Ministry of Science and Technology, Taiwan (MOST) (109-2923-E-010-001-MY2); Yen Tjing Ling Medical Foundation (CI-109-24); TPU Development Program.
**Disclosures**. The authors declare no conflicts of interest.
**Contributions.** I.M and O.M developed the idea and initiated the work. I.M., O.M. supervised the work. S.C, W.C. and C.Y. carried out the numerical analysis and I.M, O.M. and C.Y. discussed the results. I.M and O.M wrote the first draft of the manuscript and all the authors were involved in the subsequent drafts.

## References

1. Alpmann, C., Esseling, M., Rose, P. & Denz, C. Holographic optical bottle beams. *Appl. Phys. Lett.* **100,** 111101 (2012).
2. J. Arlt, M.J. Padgett, Generation of a beam with a dark focus surrounded by regions of higher intensity: the optical bottle beam. *Opt. Lett*., 25(4), 191, (2000).
3. L. Li, W. M. Lee, X. Xie, W. Krolikowski, A. V. Rode, and J. Zhou, "Shaping self-imaging bottle beams with modified quasi-Bessel beams," *Opt. Lett*. 39, 2278-2281 (2014)
4. L. Isenhower, W. Williams, A. Dally, and M. Saffman, Atom trapping in an interferometrically generated bottle beam trap, *Opt. Lett*. **34**, 1159 (2009).
5. P. Zhang, Z. Zhang, J. Prakash, S. Huang, D. Hernandez, M. Salazar, D. N. Christodoulides, and Z. Chen, Trapping and transporting aerosols with a single optical bottle beam generated by Moire techniques, *Opt. Lett*. **36**, 1491 (2011).
6. B. Melo, I. Brandao, B. Pinheiro da Silva, R.B. Rodrigues, A.Z. Khoury, and T. Guerreiro. Optical Trapping in a Dark Focus. *Phys. Rev. Appl*., **14,** 034069 (2020)
7. Z. Yang, X. Lin, H. Zhang, Y. Xu, L. Jin, Y. Zou, X. Ma, Research on the special bottle beam generated by asymmetric elliptical Gaussian beams through an axicon-lens system. *Optics and Lasers in Engineering* 2020, 126, 105899.
8. Isenhower, L., Williams, W., Dally, A. & Saffman, M. Atom trapping in an interferometrically generated bottle beam trap. *Opt. Lett.* **34,** 1159–1161 (2009).
9. B. S. Luk'yanchuk, R. Paniagua-Domínguez, I. V. Minin, O. V. Minin, and Z. Wang, "Refractive index less than two: photonic nanojets yesterday, today and tomorrow [Invited]," *Opt. Mater. Express* 7, 1820–1847 (2017).
10. Y. Zhou, H. Gao, J. Teng, X. Luo, and M. Hong, "Orbital angular momentum generation via a spiral phase microsphere," *Opt. Lett*. 43, 34–37 (2018).
11. A. A. R. Neves. Photonic nanojets in optical tweezers. Journal of Quantitative Spectroscopy and Radiative Transfer, 162, 122-132 (2015)
12. Minin I V, Minin O V, Pacheco-Peña V, M Beruete. Subwavelength, standing-wave optical trap based on photonic jets. *Quantum Electron*. **2016**, *46*, 555,
13. Minin I V, Minin O V, Cao Y, Z. Liu, Y. Geints & A. Karabchevsky. Optical vacuum cleaner by optomechanical manipulation of nanoparticles using nanostructured mesoscale dielectric cuboid. *Sci. Rep.* **2019**, *9*, 1-8,
 4. D. Lu, M. Pedroni, L. Labrador-Pez, M. I. Marqus, D. Jaque, and P. Haro-Gonzlez. Nanojet Trapping of a Single Sub-10 nm Upconverting

Nanoparticle in the Full Liquid Water Temperature Range. *Small* 2021, 17, 2006764

15. Minin I V, Minin O V. Recent trends in optical manipulation inspired by mesoscale photonics and diffraction optics. *J. Biomed. Photonics Eng.* **2020**, *6*, 020301.

16. M.-S. Kim, T. Scharf, S. Mühlig, C. Rockstuhl, and H. P. Herzig. "Generation of highly confined optical bottle beams by exploiting the photonic nanojet effect", Proc. SPIE 8274, Complex Light and Optical Forces VI, 82740U (10 February 2012);

17. Y. Zhou and M. Hong. Formation of a three-dimensional bottle beam via an engineered microsphere. *Photonics Research* 9(8), 1598 (2021)

18. Tellal, A., Ziane, O., Jradi, S., Stephan, O. & Baldeck, P. L. Quadratic phase modulation and diffraction-limited microfocusing generated by pairs of subwavelength dielectric scatterers. *Nanophotonics* 8, 1051–1061 (2019).

19. V. Pacheco-Peña, J. A. Riley, C.-Y. Liu, O. V. Minin and I. V. Minin. Diffraction limited photonic hook via scattering and diffraction of dual-dielectric structures. *Sci Rep* (2021), accepted

20. A. Taflove and S. Hagness, *Computational Electrodynamics: The Finite Difference Time Domain Method* (Artech House, 2005).

21. M. Dienerowitz, M. Mazilu, P. J. Reece, T. F. Krauss and K. Dholakia. Optical vortex trap for resonant confinement of metal nanoparticles. *Opt. Express* 16(7), 4991 (2008)

22. I. V. Minin, Yu. E. Geints, A. A. Zemlyanov, and O. V. Minin, "Specular-reflection photonic nanojet: physical basis and optical trapping application," *Opt. Express* 28, 22690-22704 (2020)

23. Zhong, M.-C., Wei, X.-B., Zhou, J.-H., Wang, Z.-Q. & Li, Y.-M. Trapping red blood cells in living animals using optical tweezers. *Nat. Commun*. 4, 1768 (2013).